\pdfoutput=1

\documentclass[%
 reprint,
 superscriptaddress,
 nobibnotes,
 amsmath,amssymb,
 aps,
 showkeys,
 longbibliography,
]{revtex4-2}

\usepackage{graphicx}
\usepackage{dcolumn}
\usepackage{bm}
\usepackage{chemformula}
\usepackage[separate-uncertainty=true,separate-uncertainty,multi-part-units=single]{siunitx}
\usepackage{hyperref}
\hypersetup{colorlinks=true,allcolors=magenta}

\begin{document}

\title{Adsorption Characteristics of Refrigerants for Thermochemical Energy Storage in Metal-Organic Frameworks}

\author{Jos\'e Manuel Vicent-Luna}
\email[Corresponding author: ]{j.vicent.luna@tue.nl}
    \affiliation{Materials Simulation \& Modelling, Department of Applied Physics, Eindhoven University of Technology, 5600 MB, Eindhoven, The Netherlands}
    \author{Azahara Luna-Triguero}
\email[Corresponding author: ]{a.luna.triguero@tue.nl}
    \affiliation{Energy Technology, Department of Mechanical Engineering, Eindhoven University of Technology, 5600 MB, Eindhoven, The Netherlands}
     \affiliation{Eindhoven Institute for Renewable Energy Systems (EIRES), Eindhoven University of Technology, Eindhoven 5600 MB, The Netherlands}

\date{\today}

\begin{abstract}

The adsorption of fluorocarbons has gained significant importance as its use as refrigerants in energy storage applications. In this context, the adsorption behavior of two low global warming potential refrigerants, R125 fluorocarbon and its hydrocarbon analog, R170, within four nanoporous materials, namely MIL-101, Cu-BTC, ZIF-8, and UiO-66 has been investigated. By analyzing the validity of our models against experimental observations, we ensure the reliability of our molecular simulations. Our analysis encompasses a range of crucial parameters, including adsorption isotherms, enthalpy of adsorption, and energy storage densities, all under varying operating conditions.We find remarkable agreement between computed and observed adsorption isotherms for R125 within MIL-101. However, to obtain similar success for the rest of the adsorbents, we need to take into account a few considerations, such as the presence of inaccessible cages in Cu-BTC, the flexibility of ZIF-8, or the defects in UiO-66.   
Transitioning to energy storage properties, we investigated various scenarios, including processes with varying adsorption and desorption conditions. Our findings underscore the dominance of MIL-101 in terms of storage densities, with R125 exhibiting superior affinity over R170. Complex mechanisms governed by changes in pressure, temperature, and desorption behavior make for complicated patterns, demanding a case-specific approach.
In summary, this study navigates the complex landscape of refrigerant adsorption in diverse nanoporous materials. It highlights the significance of operating conditions, model selection, and refrigerant and adsorbent choices for energy storage applications.

\end{abstract}


\maketitle


\section{Introduction}

Refrigerants play a critical role in modern society, with their usage spanning a wide range of applications.\cite{refrigerants-04,refrigerants-05} Since the 1830s, when Jacob Perkins developed the first-generation refrigerant, refrigerant technologies have rapidly developed, leading to the discovery of second-generation refrigerants, chlorofluorocarbon (CFC), and hydrochlorofluorocarbon (HCFC) refrigerants.\cite{refrigerants-01} Using CFCs led to the depletion of the Earth's ozone layer,\cite{refrigerants-02,refrigerants-03} leading to the development of third-generation refrigerants, such as hydrofluorocarbon (HFC) refrigerants, which have zero ozone depletion potential.\cite{refrigerants-01} The use of fluorocarbon refrigerants for energy storage applications holds promise due to their favorable thermodynamic properties, environmental compatibility, high energy density, and compatibility with adsorption materials.\cite{refrigerants-review-01,refrigerants-06} 

A few compounds have been proposed in the literature for refrigerant alternatives. For example, R125 (pentafluoroethane) is a HFC refrigerant that is commonly used as a replacement for CFCs and HCFCs in refrigeration and air conditioning systems.\cite{R125-AC,R125-mix} R125, an HFC refrigerant with a low global warming potential (GWP) of 3, presents a promising environmentally friendly alternative in refrigeration applications. This compound's effectiveness and efficiency have been established through extensive studies, affirming its potential to meet cooling needs while minimizing environmental impact. On the other hand, R170 (ethane), a hydrocarbon refrigerant, has garnered significant attention as an alternative to synthetic and HFC refrigerants.\cite{R170-01} With its GWP of 3, R170 offers an eco-friendly choice for refrigeration and air conditioning applications. Investigative efforts have focused on its inclusion in refrigerant mixtures tailored for air conditioning systems, where R170's upright performance and efficiency characteristics have been demonstrated.\cite{R170-02} Its environmental suitability is further enhanced by its excellent performance in refrigerant mixtures designed for refrigeration at extremely low temperatures, reaching as low as -80°C.\cite{R170-02} By harnessing its advantageous properties, R170-containing mixtures contribute to the sustainability of cooling technologies while lessening their environmental footprint.

Both R125 and R170 serve as exemplars of low GWP refrigerants, embodying environmentally responsible options for diverse cooling requirements. While R125 is particularly suited for refrigeration applications and has been studied extensively in this context, R170's strengths align with air conditioning systems and its associated mixtures.\cite{R125-AC,R125-mix,R170-01,R170-02,R170-mix02} Although these refrigerants boast distinct performance attributes tailored to their respective applications, their shared low GWP nature underscores their contributions to more sustainable and energy-efficient cooling solutions. In essence, R125 and R170 offer a spectrum of possibilities in the pursuit of balancing cooling needs with environmental preservation. As researchers continue to explore and optimize these properties, fluorocarbons, hydrocarbons, as well as their mixtures are poised to play a crucial role in developing efficient and sustainable energy storage solutions. Hence, to improve existing technologies is vital to understand the physicochemical properties of these compounds for their use as refrigerants in energy storage applications. Among other technologies, the emergence of using fluorocompounds through energy-efficient adsorption-based capture and separation in porous materials is vital to control their environmental impact.

The capture of refrigerants using porous materials, and in particular, HFC, has been extensively investigated in the literature. Recently, Yancey et al.,\cite{Fluorocarbons-review} and Wanigarathna et al.,\cite{refrigerants-review-01} reviewed the capture of HFC in porous materials, such as zeolites,\cite{zeo01,zeo-02,zeo-03,zeo-04,zeo-05}, silica gel,\cite{SG-01,SG-02} activated carbons,\cite{AC-01,AC-02,AC-03,AC-04,AC-05} and metal-organic frameworks (MOFs),\cite{MOF-01,MOF-02,MOF-03,MOF-04,MOF-05,MOF-06,MOF-07,MOF-08,MOF-09} discussing their practical applications. Among the rest of porous materials, MOFs are emerging as promising adsorbents for capturing and separating fluorocompounds. MOFs exhibit excellent gas adsorption performance and specific structural, physical, and chemical properties that lead to high fluorocompound adsorption capacity.\cite{refrigerants-review-01} While the adsorption of hydrocarbons in porous materials, particularly ethane (R170), has been broadly studied in the literature, (see reviews \cite{R170_rev-01,R170_rev-02} and references therein) only a few studies focused on R125.\cite{zeo-05,R125-mix,R125-specific01,Wanigarathna-01,Wanigarathna-02,Wanigarathna-03} However, regarding refrigerants adsorption, there is a lack of connection between theoretical and experimental studies. Thus, a description of the molecular mechanisms governing the adsorption process of these refrigerants in MOFs still needs to be explored. 

In this study, we analyze the particular interactions of R125 and R170 molecules with various functional MOFs, namely MIL-101, Cu-BTC, ZIF-8, and UiO-66, and discuss the molecular mechanisms governing the adsorption process. However, the working capacity, adsorption energies, and energy storage densities are more important decisive factors than the absolute adsorption capacity since those factors decide the overall efficiency of the process. We investigated the role of the operating conditions in the capture and energy storage properties of these two refrigerants. R125 and R170 share the same molecular structure $C_2X_6$, being $X$ hydrogen or fluorine atoms. This substitution of hydrogen atoms by the more interacting fluorine atoms drastically changes the physicochemical properties of the molecules, in particular, their polar nature. These changes will impact their adsorption mechanisms in MOFs, and understanding these mechanisms at a molecular level is critical for improving the existing refrigeration technologies. For the successful industrial-scale application of MOFs for fluorocompound capture, further research is required to design high-performance MOFs with optimum adsorption capacity for fluorocompound capture for energy storage applications.

\section{Methodology}

We have studied the adsorption and energy storage properties of R125 and R170 refrigerants in MIL-101, Cu-BTC, ZIF-8, and UiO-66. First, using the RASPA software,\cite{RASPA,Ariana-RASPA} we computed the adsorption isotherms by means of classical Monte Carlo simulations in the Grand Canonical ensemble. We have used a combination of Lennard-Jones and Coulombic potentials to describe the non-bonded interactions between molecules and adsorbents. The Lennard-Jones parameters that describe the MOFs interactions were taken from DREIDING force field,\cite{DREIDING} except for the metal atoms, which were taken from the UFF force field,\cite{UFF}, which is a common choice to describe gas adsorption in MOFs. The van der Waals interactions were truncated using a cut-off of 12 Å, and the cross interactions were computed using the Lorentz-Berthelot mixing rules.\cite{LB} The electrostatic interactions were computed using the Ewald summation method.\cite{Ewald}  

The initial structures of the studied MOFs are available in the RASPA repository.\cite{RASPA} However, not all the frameworks included the partial charges to describe the long-range electrostatic interactions. While models for MIL-101 and Cu-BTC have been previously reported in the literature,\cite{qMIL,qCu-BTC} we have used the charge equilibration method EQeq \cite{EQeq} to estimate the partial charges of UiO-66 and ZIF-8 derived MOFs. To account for the flexibility of ZIF-8, we have used two frameworks, as reported by Fairen et al.,\cite{Fairen-ZIF-8} where the imidazole ligands are oriented in different directions. This is a typical response of this adsorbent to the interaction with specific molecules. UiO-66 is a complex porous material that can hold a variety of crystal defects, commonly in the form of missing organic ligands.\cite{UiO-Gab,UiO-def} Because of its high symmetry, we randomly removed different organic ligands to model the defective structure of UiO-66, adding OH groups to compensate for the loose of atom connections. The crystallographic data of the studied MOFs, including the partial charges, are provided in the Supporting Information.

The definition of R125 was adopted and modified from the full-atom model reported by Peguin et al.,\cite{R125-model} for its analog R134a, which was derived from the generic OPLS-AA force field.\cite{OPLS} For R170, we have compared the performance of two models for describing its adsorption in MOFs. On the one hand, we have used a full-atom model based on the OPLS-AA force field, similar to the one for R125. On the other hand, we have tested the pseudo-atom model included in the TraPPE force field.\cite{TraPPE1,TraPPE2} The complete description of the molecules and force field are also provided in the Supporting Information.

The GCMC simulations provided the absolute loading capacity at specific pressure and temperature conditions. However, to compare the performance of different adsorbents, it is convenient to analyze the working capacity ($\Delta W$), also referred to as release capacity. This capacity is the loading difference between adsorption and desorption conditions, i.e., the loading variation when modifying the pressure, temperature, or both simultaneously:

\begin{equation} \label{eq:dW}
    \Delta W (\Delta P, \Delta T) = q_{ads}(P_{ads},T_{ads}) - q_{des}(P_{des},T_{des}) 
\end{equation}

\noindent where q$_{ads/des}$, P$_{ads/des}$ and T$_{ads/des}$ are the loading, pressure, and temperature at specific adsorption and desorption conditions, respectively.  Finally, the variations in pressure and temperature are denoted as: $\Delta P = P_{ads}-P_{des}$ and $\Delta T = T_{ads}-T_{des}$ 

In addition to the adsorption properties, we computed the specific enthalpy of adsorption or simply adsorption enthalpy through the GCMC simulations using the fluctuation method.\cite{fluctuations} Through this property, we can estimate the amount of heat released or required during adsorption/desorption cycles. This is so-called the thermochemical storage density (SD), which is calculated by the integration of the isosteric enthalpy of adsorption ($\Delta h(q)$) between adsorption and desorption conditions\cite{Lehmann_Assessment,HT-ACs,HT-Zeo}:

\begin{equation} \label{eq:SD}
    SD = \int_{q_{ads}}^{q_{des}} \Delta h(q)dq 
\end{equation}

To obtain the release capacity under varying conditions and to integrate eq. \ref{eq:SD}, it is convenient to employ smooth functions for the adsorption isotherms and enthalpy of adsorption curves. The latter were interpolated using cubic splines. We used the fitting module of RUPTURA software \cite{RUPTURA} to fit the adsorption isotherms to a Langmuir-Freundlich dual-site ($i$=2) isotherm model:

\begin{equation} \label{eq:LF}
    q(P) = \sum_{i} q_i^{sat} \frac{b_iP^{\nu_i}}{1+b_iP^{\nu_i}}
\end{equation}

\section{Results and Discussion}

Compared to investigations on light hydrocarbons, the exploration of HFCs has received less attention in the literature, particularly in terms of validating simulations against experimental observations. To assess the models' capability to describe the adsorption behaviors of the chosen refrigerants within porous materials, we have compared computed adsorption isotherms and data reported in the existing literature.\cite{Wanigarathna-01,Wanigarathna-thesis} The results of this comparison are presented in Figure 1, illustrating the adsorption of R125 in MIL-101, Cu-BTC, ZIF-8, and UiO-66 adsorbents. MIL-101, a MOF distinguished by its relatively larger cavities, presents a complex crystalline structure with three primary cavities of about 6, 25, and 33 Å as deduced from the pore size distribution (PSD) analysis. In Figure 1a), a schematic representation of MIL-101 is accompanied by the experimental and computed adsorption isotherms of R125 at 293 K. Even accounting with the structural complexity of this MOF, a reasonable agreement between the simulated and experimental outcomes validates the selected models for both the adsorbent and the HFC refrigerant. Notably, among the range of adsorbents studied, MIL-101 stands out for its remarkable capacity to adsorb R125, demonstrating an exceptional level of almost 20 mol/kg at a pressure of 10 bar.

\begin{figure*}[!t]
    \centering
    \includegraphics[width=\textwidth]{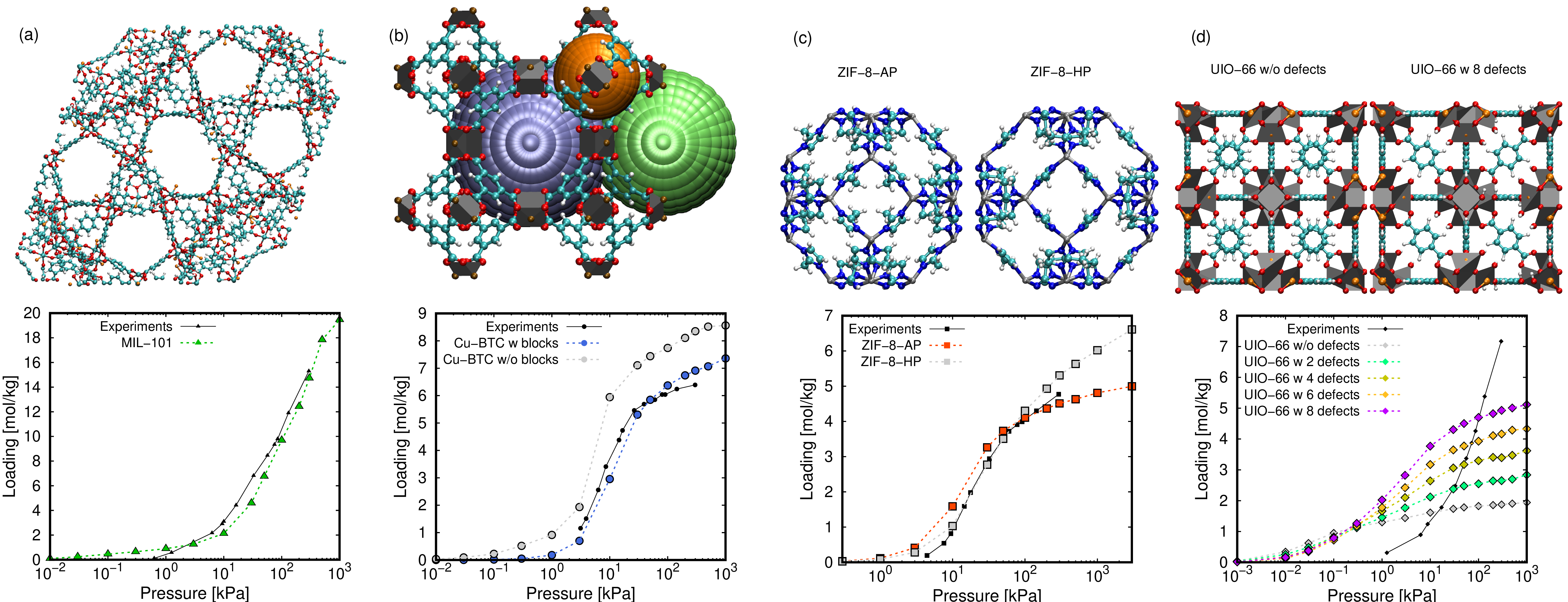}\\
    \caption{Representative snapshot of each MOF and adsorption comparison between experimental \cite{Wanigarathna-01,Wanigarathna-thesis} and computed isotherms or R125 in a) MIL-101, b) Cu-BTC, c) ZIF-8, and d) UiO-66 at 293 K.}
    \label{fig:Fig01}
\end{figure*}

The second adsorbent under investigation is Cu-BTC, another prominent MOF characterized by a simpler crystalline structure and pore network than MIL-101. However, in our initial attempt to replicate the experimental findings, we encountered a consistent deviation in the computed results, which led to an overestimation of the adsorbed quantity (see Figure 1b). Like MIL-101, Cu-BTC also contains three distinct cavity types (as depicted in Figure 1b) – measuring 5.5, 10, and 12 Å – interconnected through narrow windows, cavities that have been extensively explored in previous research \cite{Cu-BTC_JJ,Cu-BTC-ILj,Cu-BTC_Aza}. Among these, the small cavities, referred to as T1, are linked to the largest cavity (L3) containing open metal sites, while remaining detached from the intermediate cage, L2. Given that the kinetic diameter of R125 refrigerant is around 4.4 Å, a size comparable to the triangular windows granting access to the small cages, our findings propose that R125 cannot access these tiny cavities. Consequently, R125 molecules solely adsorb within the L2 and L3 cages of Cu-BTC. To corroborate this idea, additional simulations were conducted where R125 molecules were prevented from entering the small cages of Cu-BTC. Employing this approach, our simulated outcomes aligned with the experimental curve across all pressure values, reaffirming the model's validity.

With the next adsorbent, our focus shifted to ZIF-8, a nanoporous material well-known for its structural flexibility. Previous studies have demonstrated the induction of structural alterations in ZIF-8 by various molecules.\cite{Fairen-ZIF-8} A prevalent mechanism involves the rotation of imidazole ligands, which constitute the primary cavity of the MOF, measuring approximately 11 Å in size. Conducting adsorption calculations within flexible frameworks poses a formidable challenge, requiring approximations to capture these complex molecular mechanisms. A common strategy involves employing distinct crystal structures as rigid frameworks to simplify the process. As depicted in Figure 1c), two variants of ZIF-8 – labeled ZIF-8\_AP and ZIF-8\_HP – are showcased, as introduced by Fairen et al.,\cite{Fairen-ZIF-8} with differing orientations of imidazole ligands. Notably, their respective adsorption isotherms converge with experimental values. The outcomes indicate that R125 initially adsorbs in the ZIF-8\_HP configuration, displaying excellent agreement between experimental and computed data. However, as pressures ascend to higher values, the computed results for ZIF-8\_HP begin to overestimate the experimental data. In contrast, ZIF-8\_AP overestimates R125 adsorption at lower pressures, converging toward the experimental curve with increasing pressure. These findings hint at a potential structural transformation of ZIF-8 upon R125 adsorption.

The last MOF chosen for R125 capture is UiO-66, a robust framework marked by high network connectivity and cavities measuring approximately 6-7 Å. UiO-66 shows the highest framework density alongside the lowest pore size and surface area among the four selected adsorbents. Surprisingly, despite these characteristics, experimental data indicates that UiO-66 exhibits a superior adsorption capacity compared to Cu-BTC and ZIF-8, with an unusual uptake of over 7 mol/kg at 200 kPa and 293 K. However, our computations predict a substantially lower adsorption quantity of 2 mol/kg under identical conditions, and notably, the onset of adsorption pressure commences several orders of magnitude earlier than the experimental observations (Figure 4d). This considerable discrepancy might potentially arise from crystal defects present in the experimental sample, which could simultaneously decrease framework density while increasing the pore size and volume within the adsorbent. These defects, typically in the form of missing organic ligands, have been extensively documented in existing literature.\cite{UiO-def,UiO-Gab} To understand the disparity between experimental and simulated adsorption isotherms, we introduced different concentrations of defects into the UiO-66 model, as outlined in the methodology. In Figure 4d), a schematic depiction of full UiO-66 contrasts with a sample incorporating structural defects, illustrating their influence on the adsorption isotherm. As anticipated, elevating defect concentration enhances R125 adsorption, predominantly manifesting at a pressure of 1 kPa, while the low coverage regime remains relatively unaffected by these defects. Despite these efforts, the computed isotherm does not align with experimental measurements even at the highest examined defect concentration, i.e., eight missing ligands per unit cell. 

Although one could continue incorporating defects to elevate adsorbed quantities, a higher defect concentration may trigger significant structural shifts unaccounted for in the models. Consequently, an exploration of the adsorbent's stability becomes crucial. Alternatively, other types of crystal defects in the experimental sample could also influence the isotherm shape and saturation capacity. Furthermore, the presence of defects might substantially alter the structure of UiO-66 during R125 adsorption, potentially resulting in adsorption occurring on the external surface rather than within the micropores. Given these multifaceted considerations, coupled with the lack of supplementary experimental data to corroborate the findings of Wanigarathna et al.,\cite{Wanigarathna-01,Wanigarathna-thesis} and recognizing that structural stability lies beyond the scope of this study, we have excluded UiO-66 from subsequent analyses.

\begin{figure}[!h]
    \centering
    \includegraphics[width=0.4\textwidth]{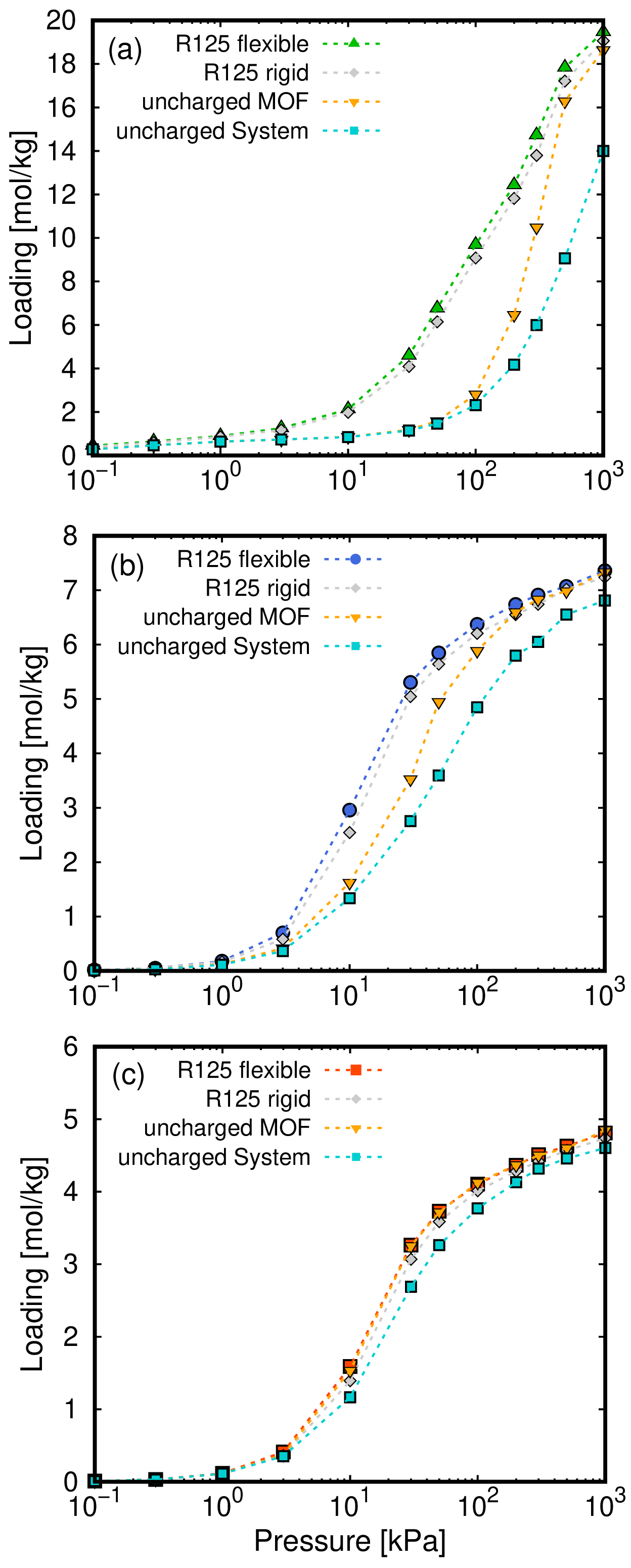}\\
    \caption{Adsorption isotherms of R125 in a) MIL-101, b) Cu-BTC, and c) ZIF-8 at 293 K, using different model simplifications.}
    \label{fig:Fig02}
\end{figure}

\begin{figure}[!t]
    \centering
    \includegraphics[width=0.4\textwidth]{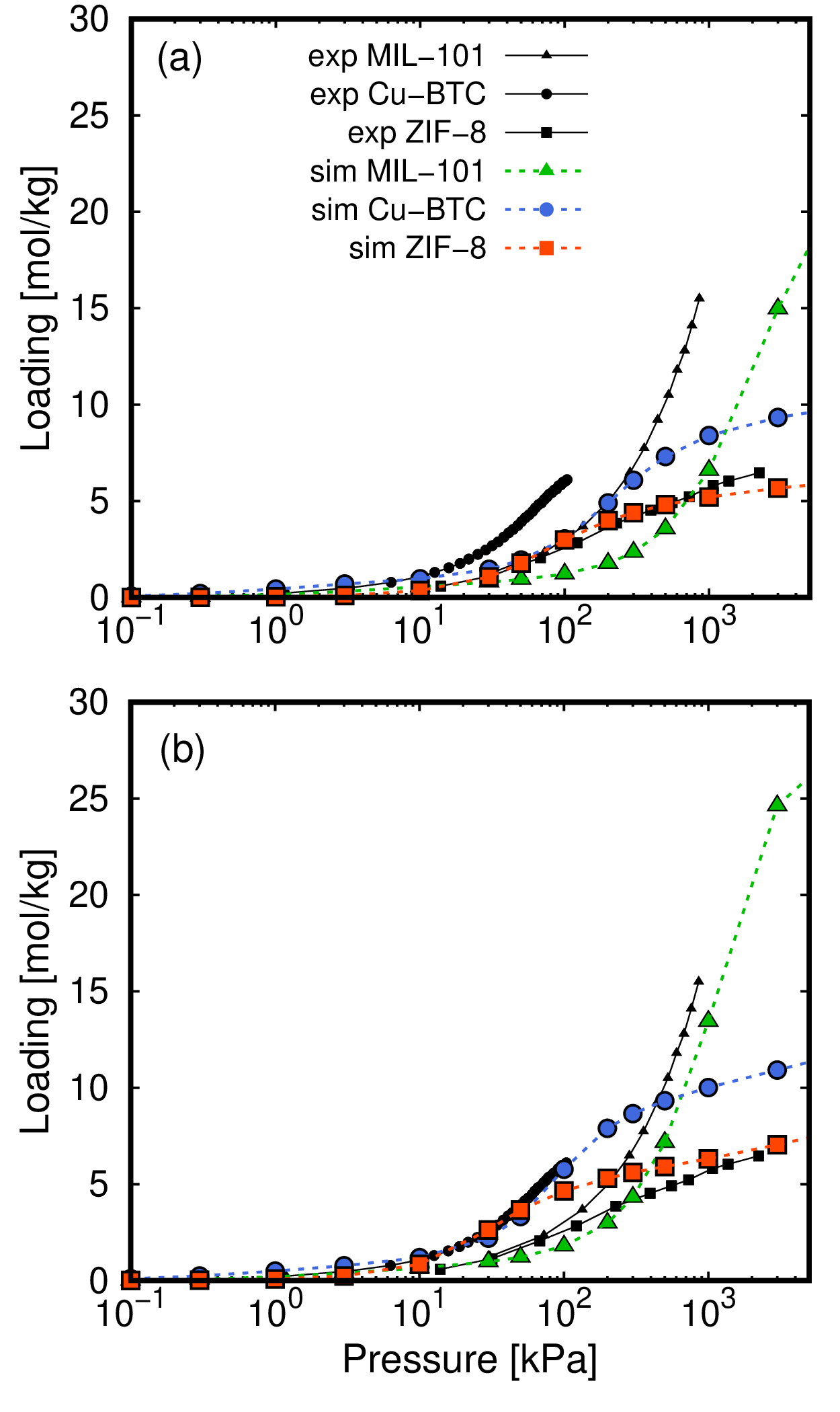}\\
    \caption{Comparison between experimental \cite{exp-MIL-101,exp-Cu-BTC,Fairen-ZIF-8} and computed adsorption isotherms of R170 in MIL-101 at 298 K, in Cu-BTC at 296 K, and in ZIF-8 at 273 K, using a), the full-atom model (OPLS-AA) and b), the pseudo-atom model (TraPPE).}
    \label{fig:Fig03}
\end{figure}

The outcomes depicted in Figure 1 corroborate the validity of the selected models in describing R125 adsorption within MOFs, provided that due considerations are taken into account. The balance between accuracy in portraying targeted properties and minimizing computational cost is essential in computational studies. Consequently, a rational model simplification, retaining predictive power, holds immense value for extensive computations necessary in examining elaborate processes or screening diverse materials. Our investigation delved into the impact of three model simplifications: i) regarding the R125 compound as a rigid body instead of a fully flexible molecule, ii) omitting Coulombic interactions between MOFs and the molecule, and iii) eliminating all Coulombic interactions within the system. These simplifications appreciably shorten the number of interactions during the simulations, influencing computational costs. Figure 2 compares R125 adsorption isotherms in MIL-101, Cu-BTC, and ZIF-8, integrating these model simplifications. Curiously, all three simplifications yield reasonable outcomes in ZIF-8, with slight underestimations in loading when all Coulombic interactions are removed. However, charged models significantly control R125 adsorption in MIL-101 and Cu-BTC. Nonetheless, deploying rigid molecules yields precise results across all examples comparable to the full-flexible model. A marginal underestimation of loading emerges in the rigid R125 model, aligning well with the original isotherm's form and adsorbed quantity. These findings affirm the validity of the rigid approach for extensive simulations, enabling the acquisition of thermodynamic properties of R125 confined within MOF pores on a large scale. However, we must emphasize that the impact of model rigidity on dynamic properties and its behavior in the liquid phase still needs to be tested, requiring validation before its application.

\begin{figure*}[!t]
    \centering
    \includegraphics[width=0.8\textwidth]{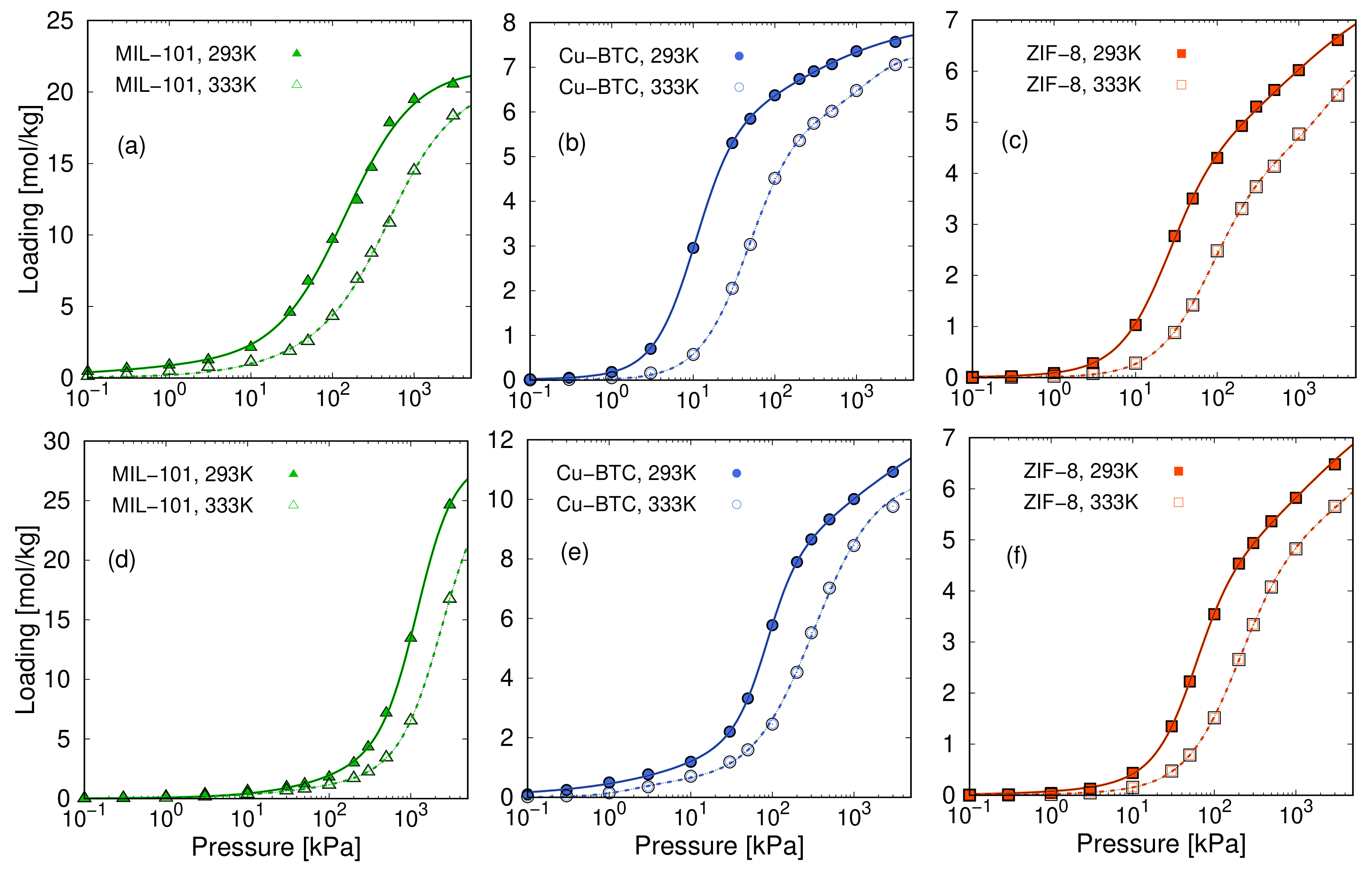}\\
    \caption{ Adsorption isotherms of a), b), and c) R125 and d), e), and f) R170, in MIL-101, Cu-BTC, and ZIF-8 at 293 K and 333 K. Symbols correspond to computed values using GCMC simulations and lines are fitting to the isotherm model.}
    \label{fig:Fig04}
\end{figure*}

Having comprehensively investigated R125's adsorption within MOFs, our attention turns to contrasting its performance with R170 refrigerant for heat storage applications. R170, commonly known as ethane, serves as a common refrigerant. In contrast to other refrigerants, adsorption behaviors of light hydrocarbons have been extensively examined in the literature, yielding an array of available models. Nevertheless, not all these models prove apt for studying hydrocarbon adsorption on porous materials. In this context, our study employed two distinct models for R170: a full-atom model akin to the R125 model and the pseudo-atom model found within the TraPPE forcefield. Figure 3 offers a comparison between these two models and experimental R170 adsorption in MIL-101, Cu-BTC, and ZIF-8. Our findings reveal that the full-atom model tends to underestimate adsorption in MIL-101 and Cu-BTC, manifesting reasonable alignment solely with ZIF-8's simulated-experimental results. Conversely, the TraPPE pseudo-atom model yields a fair concurrence for all three MOFs, with a satisfactory agreement in Cu-BTC and notable but manageable discrepancies in MIL-101 and ZIF-8. It is worth noting that, unlike R125, R170 can enter into the small cavities of Cu-BTC, as corroborated by the agreement between experimental and computed isotherms, and documented in existing literature.\cite{Cu-BTC_JJ}

Our exploration of thermochemical energy storage properties is based on adsorption and desorption cycles, simultaneously considering variations in pressure and temperature conditions. Figure 4 gathers the adsorption isotherms of both R125 and R170 within MIL-101, Cu-BTC, and ZIF-8, at temperatures of 293 K and 333 K. To process these curves, we fit the computed isotherms to a dual-site Langmuir Freundlich isotherm model, with the corresponding parameters listed in Table 1. After delving into the adsorption characteristics of these working pairs, we analyzed the performance of these processes under diverse operational conditions, involving adsorption and desorption pressures (P$_{ads}$ and P$_{des}$) and temperatures (T$_{ads}$ and T$_{des}$). As reference values, we maintained the adsorption pressure at P$_{ads}$ = 200 kPa, with adsorption and desorption temperatures fixed at T$_{ads}$ = 293 K and T$_{des}$ = 333 K, respectively. In this context, we quantified the release capacity for each MOF-adsorbate pair, simultaneously adjusting the temperature and pressure throughout the adsorption/desorption cycles. To obtain a comprehensive overview of the impact of these operating conditions, we assessed three distinct scenarios: i) Maintaining P$_{ads}$ while decreasing P$_{des}$ for T$_{ads}$ = T$_{des}$, ii) keeping T$_{ads}$ at 293 K and T$_{des}$ at 333 K while setting P$_{ads}$ equal to P$_{des}$, and iii) fixing T$_{ads}$ at 293 K, T$_{des}$ at 333 K, and fixing P$_{ads}$ while lowering P$_{des}$. These scenarios mirror pressure-swing adsorption (PSA), temperature-swing adsorption (TSA), and pressure-temperature swing adsorption (PTSA) processes, respectively.

\begin{table*}[!ht]
    \centering
    \caption{Parameters of the dual-site Langmuir-Freundlich isotherm model for the adsorption of R125 and R170 at 293 K and 333 K in MIL-101, Cu-BTC, and ZIF-8. $q_{sat}$ is given in mol/kg, $b$ in kPa$^{-1}$, and $\nu$ is dimensionless.}
    \begin{tabular}{|l|l|l|l|l|l|l|l|l|}
    \hline
        R125 & \multicolumn{4}{c|}{293 K} & \multicolumn{4}{c|}{333 K} \\ \hline
        ~ & site & q$_{ads}$ & b & $\nu$ & site & q$_{ads}$ & b & $\nu$ \\ \hline 
        MIL-101 & 1 & 1.47412 & 1.05509 & 0.474721 & 1 & 13.66 & 0.015586 & 0.687025 \\
        ~ & 2 & 20.2103 & 0.00609846 & 1.024 & 2 & 7.83051 & 0.000126077 & 1.44533 \\ \hline
        Cu-BTC & 1 & 2.56266 & 0.0359348 & 0.632713 & 1 & 1.27658 & 4.11684e-06 & 1.7161 \\ 
        ~ & 2 & 5.44565 & 0.0162049 & 1.75207 & 2 & 6.07579 & 0.00413181 & 1.41063 \\ \hline
        ZIF-8 & 1 & 4.01341 & 0.0088463 & 1.47029 & 1 & 4.48204 & 0.00352973 & 1.25497 \\ 
        ~ & 2 & 3.89465 & 0.0152228 & 0.618818 & 2 & 1.99951 & 9.97486e-06 & 1.47481 \\ \hline
        R170 & \multicolumn{4}{c|}{~} & \multicolumn{4}{c|}{~} \\ \hline
        MIL-101 & 1 & 22.5532 & 7.62306e-07 & 1.98606 & 1 & 24.2143 & 2.03385e-06 & 1.68275 \\ 
        ~ & 2 & 6.33784 & 0.0132412 & 0.73601 & 2 & 2.94606 & 0.0302716 & 0.631608 \\ \hline
        Cu-BTC & 1 & 5.37097 & 3.68737e-05 & 2.2971 & 1 & 0.500097 & 0.336992 & 1.60332 \\ 
        ~ & 2 & 8.28741 & 0.0584847 & 0.447585 & 2 & 10.2633 & 0.00136971 & 1.14042 \\ \hline
        ZIF-8 & 1 & 3.71105 & 0.00066986 & 1.79981 & 1 & 5.66035 & 0.00381952 & 0.517259 \\ 
        ~ & 2 & 4.62521 & 0.0163971 & 0.572956 & 2 & 4.65319 & 0.000823884 & 1.33983 \\ \hline
    \end{tabular}
\end{table*}

\begin{figure*}[!t]
    \centering
    \includegraphics[width=0.8\textwidth]{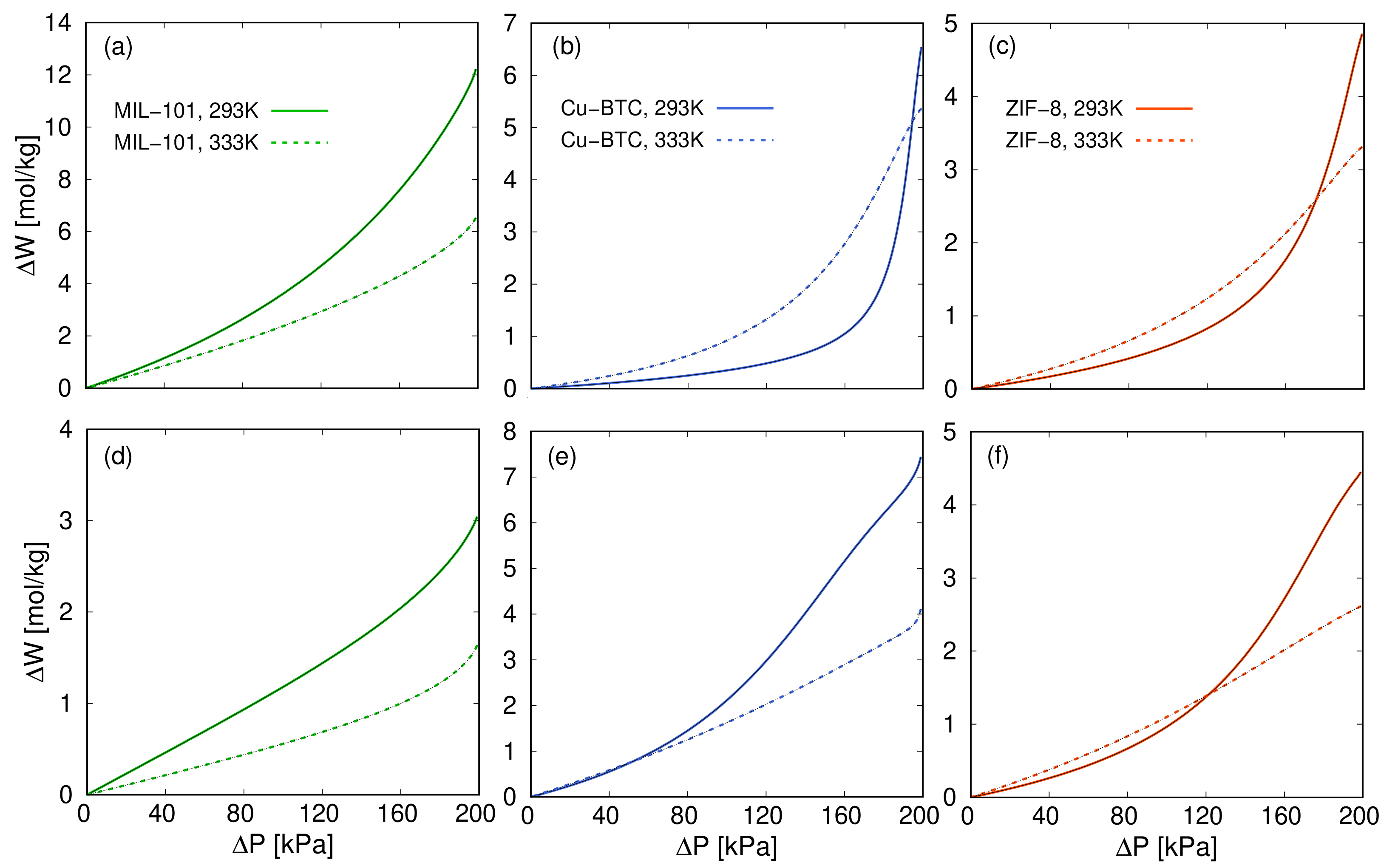}\\
    \caption{Release capacity ($\Delta$W) as a function of the pressure lift ($\Delta$P) of a), b), and c) R125 and d), e), and f) R170, in MIL-101, Cu-BTC, and ZIF-8 at 293 K and 333 K. P$_{ads}$ was fixed to 200 K and T$_{ads}$ = T$_{des}$.}
    \label{fig:Fig05}
\end{figure*}

Adsorption isotherms and isobars elucidate the relationship between loading and pressure, as well as temperature, respectively. However, it often proves more convenient to evaluate loading in terms of the difference between adsorption and desorption amounts as pressure or temperature varies. Notably, our focus encompassed both adsorption and desorption conditions as integral components of thermodynamic cycles for energy storage applications. It is worth mentioning that we did not differentiate between adsorption and desorption isotherms in this context. Taking this into account, we computed the release capacity (eq. \ref{eq:dW}) from the isotherms illustrated in Figure 4. Figures 5 and 6 depict the release capacity of R125 and R170 for the three selected MOFs across each of the scenarios previously outlined. Figure 5 compiles the release capacity under isothermal conditions of 293 K and 333 K, while varying the pressure lift ($\Delta$P). Although the adsorption isotherms exhibit similar trends across all instances (see Figure 4), the release capacity showcases distinct behavior for each adsorbent, adsorbate, and operational conditions.

\begin{figure}[!t]
    \centering
    \includegraphics[width=0.48\textwidth]{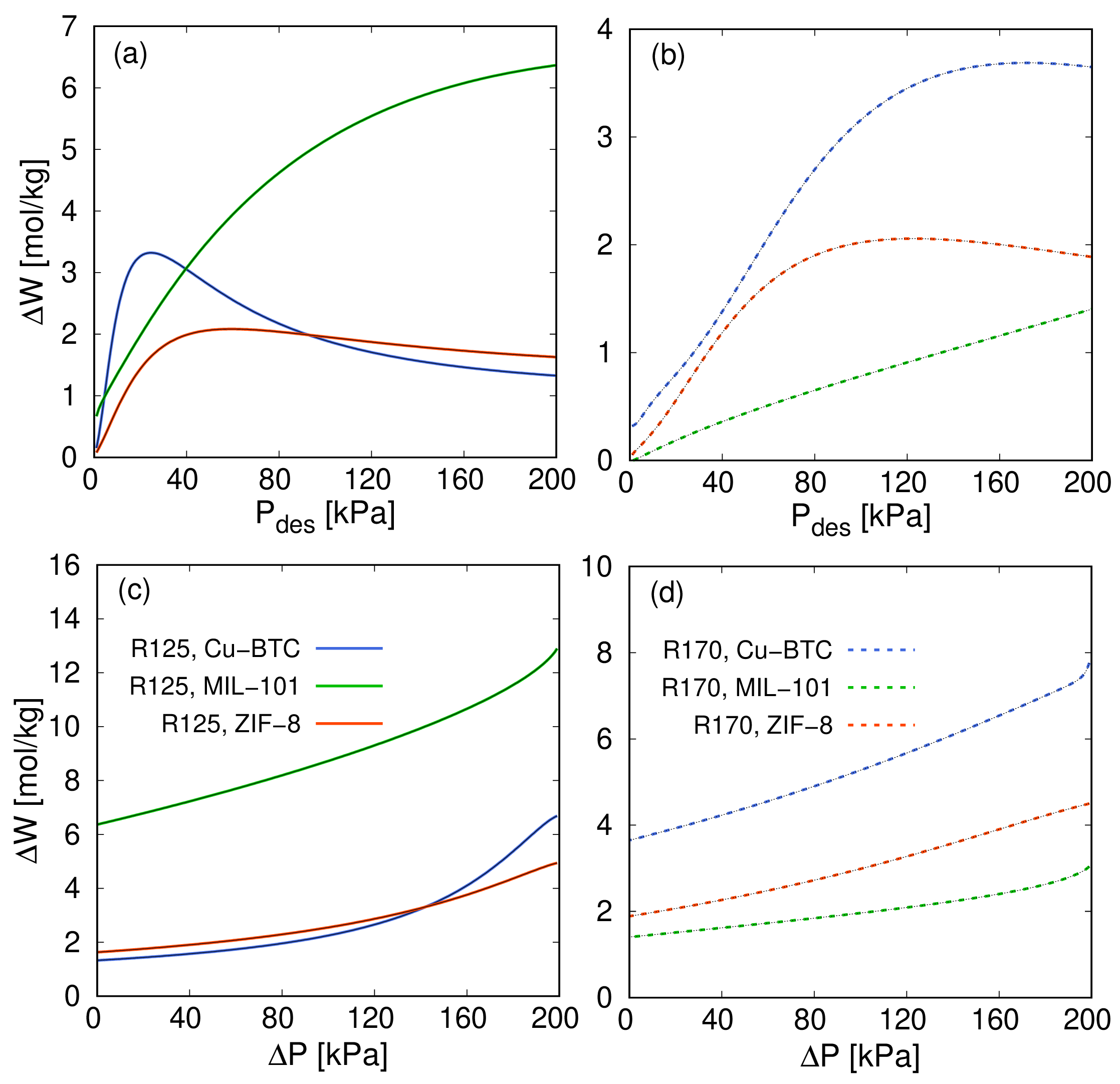}\\
    \caption{Release capacity ($\Delta$W) as a function of the desorption pressure (P$_{des}$) a) and b), and pressure lift ($\Delta$P) c) and d) for R125 (solid lines) and  R170 (dashed lines), in MIL-101, Cu-BTC, and ZIF-8. The pressure conditions for a) and b) are P$_{ads}$ = P$_{des}$ and for c) and d) are P$_{ads}$ = 200 kPa, while P$_{des}$ varies. In all cases, T$_{ads}$ = 293 K and T$_{des}$ = 333 K.}
    \label{fig:Fig06}
\end{figure}

For MIL-101, the release capacity of both R125 and R170 is higher at the lower temperature, revealing similar patterns for both refrigerants. However, R125's release capacity under these conditions considerably outperforms the capacity of R170. At a pressure lift near 200 kPa and 293 K, R125 demonstrates a release capacity of 12 mol/kg, while R170 achieves 3 mol/kg. Turning to Cu-BTC and ZIF-8, the release capacity profiles of R125 and R170 exhibit divergent tendencies. In these two materials, R125 presents a low release capacity at varying pressures, which grows significantly near a pressure lift of 200 kPa. Unlike MIL-101, the release capacity of R125 in Cu-BTC and ZIF-8 is somewhat higher at the higher temperature, outdone only by the curve for the lowest temperature, reaching the maximum pressure lift. A different dynamic emerges for R170 in these materials, with the release capacity demonstrating comparable values at both temperatures, with variations emerging at intermediate pressure lifts (around 100 kPa) in favor of the isothermal process at 293 K. Notably, unlike MIL-101, Cu-BTC, and ZIF-8 exhibit a near-equivalent release capacity, diverging by approximately 1 mol/kg between R125 and R170 across the range of 4-7 mol/kg.

Figure 6 illustrates the release capacity of both R125 and R170 within the three MOFs, encompassing the other two scenarios: an isobaric process involving a temperature shift from T$_{ads}$ = 293 K to T$_{des}$ = 333 K, and a pressure-temperature swing process concurrently altering the pressure lift across temperatures from T$_{ads}$ = 293 K to T$_{des}$ = 333 K. Similar to the prior scenario, MIL-101 boasts the highest release capacities for R125 and the lowest for R170, contrasting with Cu-BTC and ZIF-8. In the context of the isobaric process, we note a lack of consistent escalation in the release capacity as P$_{des}$ rises in Cu-BTC and ZIF-8. For instance, the release capacity of R125 demonstrates a peak at desorption pressures ranging from 30 kPa to 50 kPa for Cu-BTC and ZIF-8. This peak is either minimal or appears at significantly higher desorption pressures for R170. These intricate variations necessitate careful consideration, and the formulation of general guidelines becomes complex, thereby demanding an exhaustive evaluation of diverse working pairs and operational conditions.

\begin{figure}[!t]
    \centering
    \includegraphics[width=0.4\textwidth]{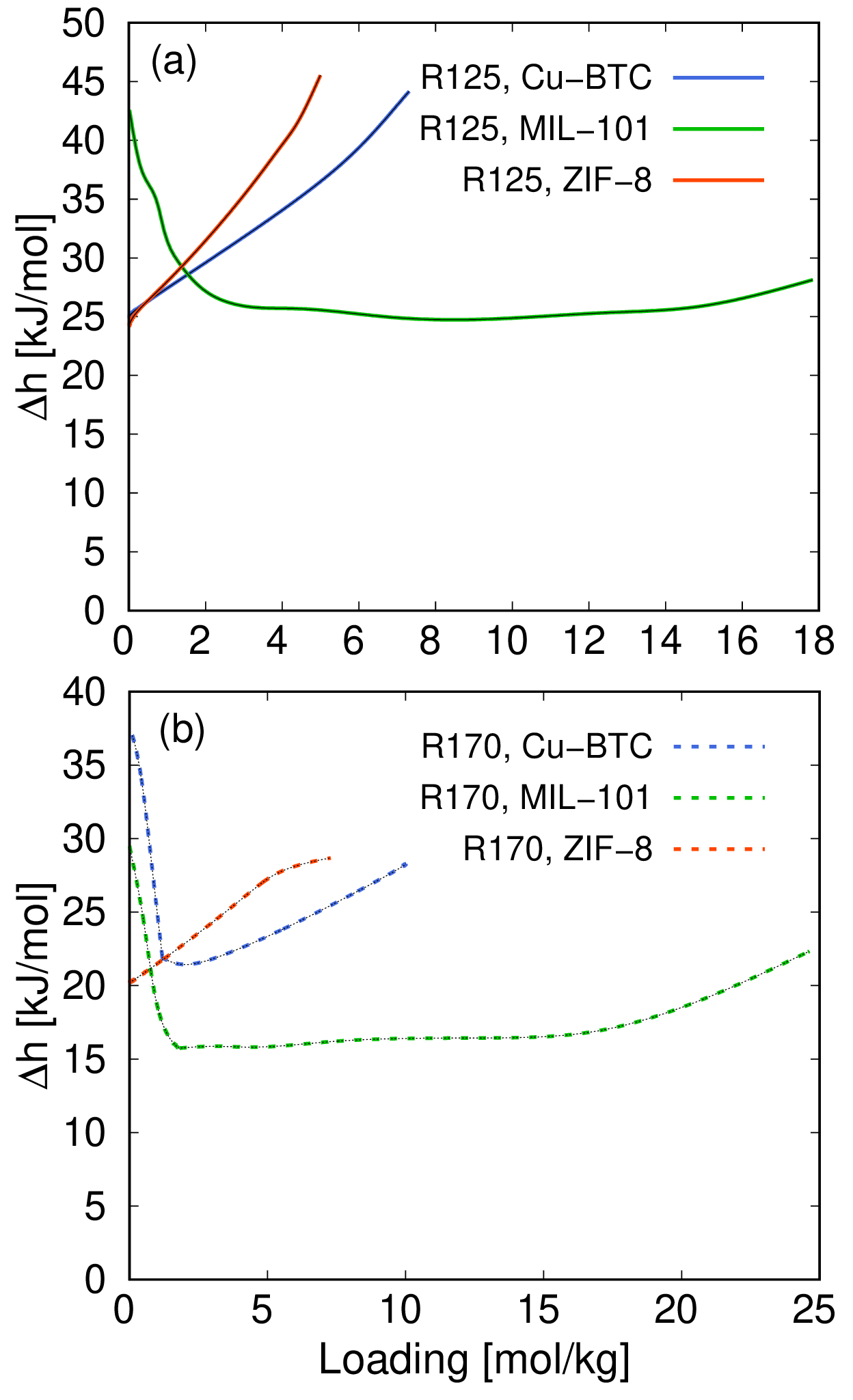}\\
    \caption{Enthalphy of adsorption as a function of the loading ($\Delta$h(q)) of a) R125, and b) R170 in MIL-101, Cu-BTC, and ZIF-8.}
    \label{fig:Fig07}
\end{figure}

Figure 7 offers insight into the loading-dependent specific enthalpy of adsorption, commonly referred to as the enthalpy of adsorption, for both R125 and R170 refrigerants across the three selected MOFs. On a broad scale, R125 exhibits greater affinity for all MOFs compared to R170, as evidenced by its higher enthalpy of adsorption. A key distinction between R125 and R170 manifests when examining their behavior at low coverage in Cu-BTC. Specifically, the enthalpy of adsorption for R170 achieves its maximum value at low loading, then diminishes significantly before rising once more as loading increases. This phenomenon arises due to R170's adsorption within the small cavities of Cu-BTC, which promotes molecular confinement and consequently leads to a higher enthalpy of adsorption.\cite{Cu-BTC_JJ} Contrarily, a similar effect is not observed in the enthalpy of adsorption for R125 in Cu-BTC, primarily because R125 only adsorbs within the larger cavities of this MOF. Nevertheless, this analogous effect materializes in the adsorption of both refrigerants in MIL-101, where the initial decrease in the enthalpy of adsorption can be attributed to the adsorption in the smaller cages of MIL-101 that are not restricted to any of the two molecules.

\begin{figure*}[!t]
    \centering
    \includegraphics[width=0.8\textwidth]{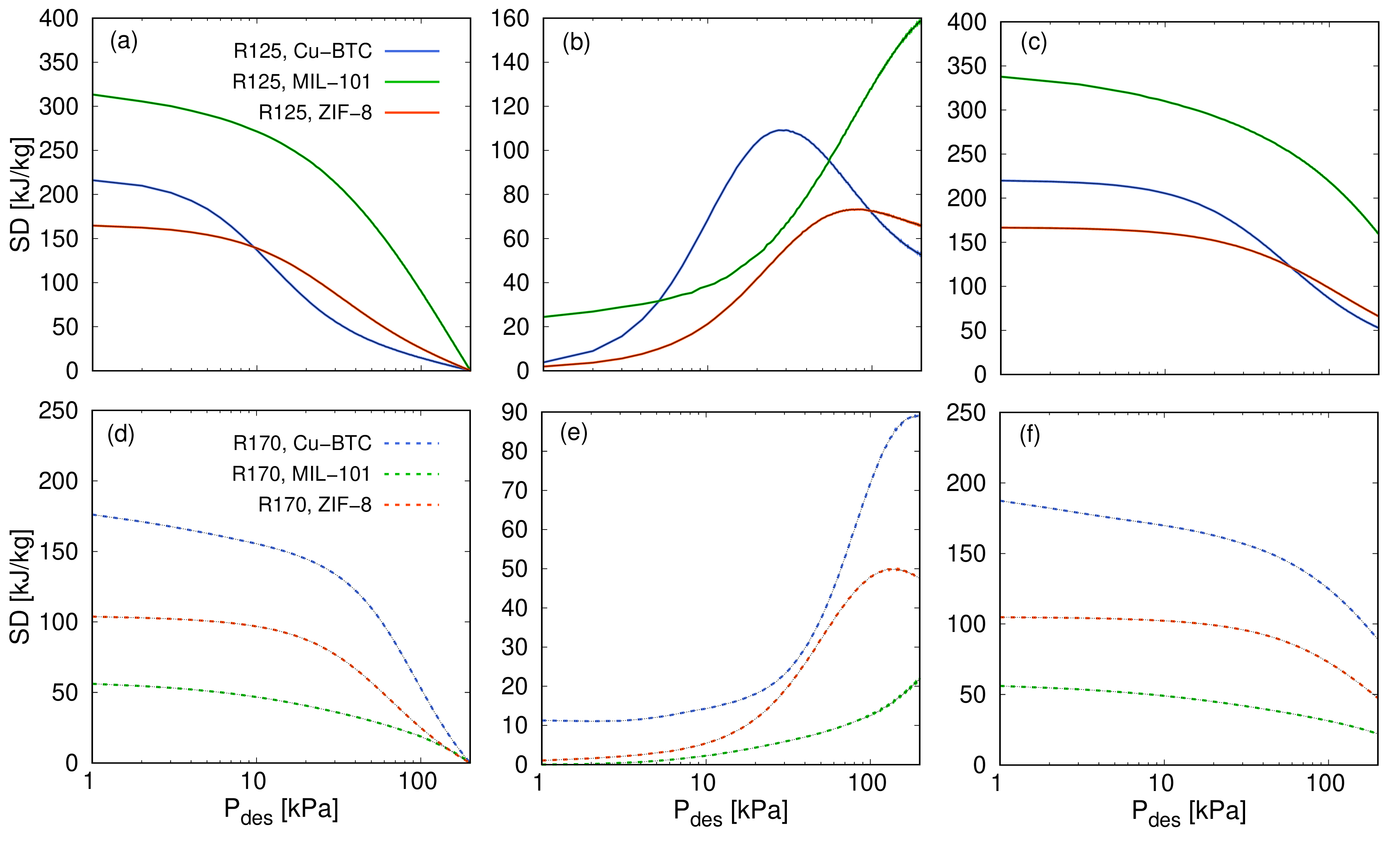}\\
    \caption{Energy storage densities (SD) of R125 (solid lines) and R170 (dashed lines) in MIL-101, Cu-BTC, and ZIF-8 as a function of the desorption pressure. The operating conditions for a) and d) are P$_{ads}$ = 200 kPa, P$_{des}$ varies, T$_{ads}$ = T$_{des}$ = 293 K; for b) and e) are P$_{ads}$ = P$_{des}$, T$_{ads}$ = 293 K, and T$_{des}$ = 333 K; and for c) and f) are P$_{ads}$ = 200 kPa, P$_{des}$ varies, T$_{ads}$ = 293 K, and T$_{des}$ = 333 K.}
    \label{fig:Fig08}
\end{figure*}

The last step of this study involved the computation of energy storage densities for each working pair, accomplished by integrating the enthalpy of adsorption curves while varying adsorption and desorption conditions. These energy storage densities serve as the critical property of an energy storage device, denoting the amount of heat that can be stored or released per unit mass or volume of adsorbent following an adsorption-desorption cycle. In alignment with the earlier Figures 5 and 6, we delve into the examination of storage densities for both R125 and R170 within the contexts of mirroring PSA, TSA, and PTSA scenarios, as depicted in Figure 8.
Across all scenarios, R125 consistently exhibits higher storage densities than R170, except for Cu-BTC, where the values are comparably close. This contrast stems from R125's confinement to larger cavities within Cu-BTC, preventing its access to the small cavities that enhance storage density. Notably, in terms of maximal values, the reduction of pressure when P$_{ads}$ = 200 kPa yields storage densities that surpass those obtained from increasing the temperature range between T$_{ads}$ = 293 K and T$_{des}$ = 333 K. This suggests that for scenarios mirroring TSA processes, a higher desorption temperature proves to be critical for maximizing storage density across all adsorbents.
Furthermore, our findings indicate minimal disparities between PSA and PTSA processes. The alterations in temperatures predominantly affect storage densities for modest pressure lifts, while negligible distinctions are noticeable for the highest storage densities achieved under fixed conditions.

In our final series of calculations, we operated under the assumption of a complete regeneration cycle, containing the entire spectrum from maximum adsorption to complete desorption conditions. To accomplish this, we integrated the enthalpy of adsorption curves (Figure 7) over the entirety of the adsorbed quantities of both R125 and R170 across the three adsorbents. The adsorption conditions are P$_{ads}$ = 200 kPa and T$_{ads}$ = 293 K, while the desorption conditions can be those combinations of P$_{des}$ and T$_{des}$ that make that q$_{des}$ approaches to zero. Using these assumptions, the maximum SD of R125 in MIL-101, Cu-BTC, and ZIF-8 are 470 kJ/kg, 246 kJ/kg, and 169 kJ/kg, respectively, while for R170 are 431 kJ/kg, 246 kJ/kg, and 183 kJ/kg, respectively. As anticipated, MIL-101 emerges as the outstanding adsorbent, boasting the highest storage densities for R170 and R125, respectively. These values double the storage density of both refrigerants when in Cu-BTC. On the other end of the spectrum, ZIF-8 demonstrates the most modest storage densities, failing to breach the 200 kJ/kg threshold for both R125 and R170. Significantly, in the studied conditions, MIL-101's storage densities for the R125 refrigerant resemble those observed in methanol adsorption within zeolites such as NaY or NaX \cite{HT-Zeo}, as well as methanol and ethanol within activated carbons \cite{HT-ACs,BPL}. These findings establish MIL-101 as an outstanding candidate for deployment within thermochemical energy storage applications.

\section{Conclusions}

In this study, we have explored the adsorption behavior of R125 and R170 refrigerants within various nanoporous materials, shedding light on the complex interplay between adsorbent properties, operating conditions, and model accuracy. Our analysis demonstrates the utility of these materials in thermochemical energy storage applications and reveals valuable insights into the performance of different refrigerants under diverse scenarios.
Through meticulous validation against experimental data, we have established the accuracy of the chosen models, enabling a precise description of adsorption phenomena. This validation process has revealed the specific requirements for each refrigerant and adsorbent combination, such as considering structural flexibility, inaccessible pores to a certain molecule, and crystal defects. The variations in behavior observed between R125 and R170 highlight the role of molecular structure and chemical composition in dictating adsorption characteristics. Considering model simplifications, we ascertain that a rigid molecule model offers a viable compromise between computational efficiency and predictive accuracy for large-scale simulations. 

Our investigation into energy storage applications underlines the significance of pressure, temperature, and adsorption-desorption cycles. The contrasting scenarios of pressure swing, temperature swing, and pressure-temperature swing processes offer a comprehensive understanding of the mechanisms governing adsorption-based energy storage. The dominance of MIL-101 in terms of energy storage densities showcases its potential as an optimal adsorbent, with R125 emerging as the refrigerant of choice due to its superior affinity and capacity for energy storage.
Our study emphasizes the complex interplay between adsorbent properties, refrigerant characteristics, and operational conditions in thermochemical energy storage applications. We have illustrated the path toward more efficient and effective energy storage solutions with an approach that combines careful model selection, experimental validation, and comprehensive scenario analysis. This work not only contributes to the fundamental understanding of adsorption phenomena but also guides the design of practical and sustainable energy storage technologies in the future.


\section*{Supporting Information}
Crystallographic data of adsorbents, molecule definition of adsorbates, and all force field parameters used in this work. At this moment, the Supporting Information is available via email to the authors upon reasonable request. The Supporting Information will be published as open access together with the manuscript once it is accepted for publication. 

\section*{Acknowledgements}

A.L-T acknowledges funding by the Irène Curie Fellowship program of the Eindhoven University of Technology, and J.M.V-L acknowledges the APSE department of TU/e for funding support and computing resources.

\section*{Author contributions}

A. L.-T. and J. M. V.-L. have contributed equally to this work.

\section*{Competing interests}

The authors declare no competing interests.

\bibliography{Arxiv_RefMOFs}













\end{document}